\documentclass[11pt]{article}

\usepackage{amssymb}
\usepackage[numbers,sort&compress]{natbib}
\usepackage{amsmath,amsfonts,graphicx,epsfig}
\usepackage{bm}
\usepackage{color}

\def\bea{\begin{eqnarray}}
\def\eea{\end{eqnarray}}
\def\be{\begin{equation}}
\def\ee{\end{equation}}
\def\ba{\begin{array}}
\def\ea{\end{array}}

\begin{document}

\setlength\arraycolsep{2pt}

\renewcommand{\theequation}{\arabic{section}.\arabic{equation}}
\setcounter{page}{1}

\begin{titlepage}

\begin{center}

\vskip 1.5 cm

{\huge \bf The string swampland constraints require multi-field inflation}

\vskip 2.0cm

{\Large 
Ana Ach\'ucarro$^{a,b}$ and Gonzalo A. Palma$^{c}$
}

\vskip 0.5cm

{\it $^{a}$Lorentz Institute for Theoretical Physics, Leiden University, 2333CA Leiden, The Netherlands \\
$^{b}$Department of Theoretical Physics, University of the Basque Country, 48080 Bilbao, Spain\\
$^{c}$Grupo de Cosmolog\'ia y Astrof\'isica Te\'orica, Departamento de F\'{i}sica, FCFM, \mbox{Universidad de Chile}, Blanco Encalada 2008, Santiago, Chile.  }

\vskip 2.5cm

\end{center}

\begin{abstract} 
An important unsolved problem that affects practically all attempts to connect string theory to cosmology and phenomenology is how to distinguish effective field theories belonging to the string landscape from those that are not consistent with a quantum theory of gravity at high energies (the ``string swampland"). It was recently proposed that potentials of the string landscape must satisfy at least two conditions, the ``swampland criteria", that severely restrict the types of cosmological dynamics they can sustain. The first criterion states that the (multi-field) effective field theory description is only valid over a field displacement $\Delta \phi \leq \Delta \sim \mathcal O(1)$ (in units where the Planck mass is 1), measured as a distance in the target space geometry. A second, more recent, criterion asserts that, whenever the potential $V$ is positive, its slope must be bounded from below, and suggests $|\nabla V| / V \geq c  \sim   \mathcal O(1)$. A recent analysis concluded that these two conditions taken together practically rule out slow-roll models of inflation. In this note we show that the two conditions rule out inflationary backgrounds that follow geodesic trajectories in field space, but not those following curved, non-geodesic, trajectories (which are parametrized by a non-vanishing bending rate $\Omega$ of the multi-field trajectory). We derive a universal lower bound on $\Omega$ (relative to the Hubble parameter $H$) as a function of $\Delta, c$ and the number of efolds $N_e$, assumed to be at least of order 60. If later studies confirm $c$ and $\Delta$ to be strictly $\mathcal O(1)$, the bound implies strong turns with $\Omega / H \geq 3 N_e \sim 180$. Slow-roll inflation in the landscape is not ruled out, but it is strongly multi-field.

\end{abstract}

\end{titlepage}

\newpage

\setcounter{equation}{0}
\section{Introduction}

A satisfactory version of cosmic inflation~\cite{Guth:1980zm, Linde:1981mu, Albrecht:1982wi, Starobinsky:1980te, Mukhanov:1981xt} within string theory is an open challenge~\cite{Baumann:2014nda}.  The stabilization of moduli, together with the realization of flat directions in the $4D$ landscape potential $V$, are two obstructions commonly encountered in the construction of de Sitter vacua~\cite{Maldacena:2000mw, Silverstein:2007ac, Covi:2008ea, Covi:2008zu, Danielsson:2009ff, Wrase:2010ew, Danielsson:2011au, Dasgupta:2014pma, Danielsson:2018ztv, Andriot:2018wzk, Dvali:2018fqu} and of working models of inflation~\cite{Brustein:1992nk, Hertzberg:2007wc, Covi:2008cn, Flauger:2008ad, Caviezel:2008tf, Caviezel:2009tu} in low energy compactifications of string theory. Attempts to understand the physical origin of the previous two obstructions have crystalized into two necessary conditions, the so-called {\em swampland criteria}~\cite{Obied:2018sgi}, that allegedly should be satisfied by $4D$ low energy effective field theories (EFT's) resulting from string theory compactifications, those belonging to the string landscape~\cite{Vafa:2005ui, Ooguri:2006in, Brennan:2017rbf}. The scalar (plus gravity) sector of these theories is typically described by an action containing several scalar fields $\phi^a$ as
\be
S = \int d^4 x \sqrt{-g}  \left[ \frac{M_{\rm Pl}^2}{2} R -  \frac{1}{2} G_{a b} (\phi)  g^{\mu \nu} \partial_\mu \phi^a \partial_\nu \phi^b - V (\phi) \right] , \label{action}
\ee
valid below some scale $\Lambda$, where $g_{\mu \nu}$ is the space-time metric, $g = \det(g_{\mu\nu})$, $R$ is the Ricci scalar, and $G_{a b}$ is a sigma model metric characterizing the geometry of the scalar field target space. The first criterion imposes a maximum distance in field space beyond which the low energy description ceases to be valid due to quantum gravity effects, even if this is not obvious from the EFT itself.  In the best understood cases,  failure of the EFT description is linked to the appearance in the string compactification of a tower of light fields, whose masses go to zero exponentially in the (proper) field distance $\Delta \phi$. The constraint states that the field range traversed by the fields must stay smaller than a given value $\Delta \sim \mathcal O (1)$~\cite{Baume:2016psm, Klaewer:2016kiy, Grimm:2018ohb, Blumenhagen:2018hsh} (in units where $M_{\rm Pl} = 1$):
\be 
\Delta \phi < \Delta.  \label{first_criterion}
\ee
The second criterion is more recent, and of a more speculative nature. It takes the form of a lower bound on the gradient of a (positive) potential $V$ in an EFT that admits a UV completion consistent with quantum gravity. It states that the shape of the landscape potential $V$ of the $4D$ effective theory is such that, for $V>0$, there will unavoidably exist steep directions for which
\be
|\nabla V| \geq c \, V ,    \label{second_criterion}
\ee
where $c$ is a positive constant of order $1$. Given that we are dealing with an EFT with several fields, in the previous expression one should take $|\nabla V| = \sqrt{G^{ab} V_a V_b}$ where $V_a = \partial V /\partial \phi^a$. The quantities $\Delta$ and $c$ parametrizing the criteria, depend on the details of the compactification leading to the low energy effective field theory. 

While their validity is under scrutiny, one may test the swampland criteria against well established phenomenological frameworks within cosmology and particle physics. The implications for cosmology emerging from these two criteria have been recently examined by the authors of Ref.~\cite{Agrawal:2018own}. In particular, they conclude that cosmic inflation is in conflict with these two criteria. The argument outlined in~\cite{Agrawal:2018own} works as follows: 

First, in single field inflation the tensor to scalar ratio is given by
\be
r = 16 \epsilon, \label{r-epsilon-single}
\ee
where $\epsilon$ is the first slow-roll parameter (see Section~\ref{first_slow_roll_par}). The background equations of motion  relate $\epsilon$ to the speed of the evolving inflaton field via $d \phi_0 / d N_e = \sqrt{2 \epsilon}$, where $N_e$ is the number of $e$folds elapsed during inflation ($e$folds and cosmic time are related via $d N = H dt$, where $H$ is the Hubble parameter during inflation). Then, since inflation must have lasted about $60$ $e$folds (from the time of horizon crossing of the largest observable scales), one concludes that $\Delta \phi \sim 60 \sqrt{2 \epsilon} = 30 \sqrt{r/2}$. Together with the first criterion~(\ref{first_criterion}), this relation implies:
\be
\Delta \phi \sim 60 \sqrt{2 \epsilon} = 30 \sqrt{r/2} < \Delta \sim \mathcal O (1) .  
\ee
Current cosmic microwave background (CMB) constraints on the presence of primordial $B$-modes require $r < 0.07$~\cite{Ade:2015tva} or, using Eq. (\ref{r-epsilon-single}), $\epsilon < 0.0044$. This implies $\Delta \phi \sim  60 \sqrt{2 \epsilon}  < 5.6$. This reveals~\cite{Lyth:1996im} that a detection of $B$-modes by the next generation of CMB experiments~\cite{Suzuki:2015zzg, Harrington:2016jrz, Abazajian:2016yjj} may lead to a tension with criterion 1. For instance, if CMB experiments observe $r = 0.01$, then $\Delta \phi \sim 60 \sqrt{2 \epsilon} = 2.1$ which would be in mild tension with $\Delta \sim \mathcal O (1)$, specially taking into account that the number of $e$folds of the duration of inflation must be greater than $60$. 

The second criterion turns out to be more severe on single field inflation. In single field inflation the background equations of motion also relate the value of $\epsilon$ to the shape of the potential driving inflation --in particular, to the gradient of the potential along the trajectory. The relation is given by:
\be
\epsilon = \frac{1}{2} \left( \frac{\nabla V}{V} \right)^2 . \label{epsilon-naive}
\ee
This relation may be used together with (\ref{second_criterion}) to infer $c < \sqrt{2 \epsilon}$. Notice that, irrespective of any observational constraints, this bound implies $60 c < \Delta$ which is in trouble with the requirement that $\Delta$ and $c$ be both of order 1. If we factor in $\epsilon < 0.0044$ (to satisfy current CMB constraints), we find $c < 0.094$.

The purpose of this note is to show that these conclusions change drastically in the case of multi-field models of inflation derived from the action (\ref{action}), because both Eqs.~(\ref{epsilon-naive}) and~(\ref{r-epsilon-single}) stop being valid. As it happens, in multi-field inflation it is {\em essential} to take into account the effects coming from the bending of the inflationary trajectory within the multi-field target space which, in string theory compactifications, is usually characterized by a non-trivial geometry. As we shall see, the effects coming from a bending trajectory in multi-field space are enough to relax the tension coming from criterion 2. On the other hand, the impact of multi-field dynamics on the bounds coming from the field range $\Delta\phi$  is more subtle, and will require sharper versions of criterion 1.  There are two competing effects here because, on a turning trajectory,   the observations give a stronger constraint on the distance traversed by the field {\em along the inflationary trajectory}, but this distance may be considerably larger than the geodesic distance that appears in the towers of light states (see \cite{Landete:2018kqf,Hebecker:2017lxm} for recent discussions of this important difference). In what follows we will ignore this distinction and make the conservative assumption that the upper bound $\Delta \sim O(1)$ applies to all field ranges, geodesic or not.  But it is an interesting caveat to the first swampland criterion that should be explored.

More to the point, in multi-field models one generically finds $r < 16 \epsilon$. For instance if fluctuations orthogonal to the inflationary trajectory are massive, with masses $M \gg H$, then they typically induce the appearance of a sound speed for the propagation of the curvature perturbation~\cite{Tolley:2009fg, Achucarro:2010jv, Achucarro:2010da} given by $c_s = (1 + 4 \Omega^2 / M^2)^{-1/2}$, where  $\Omega$ is the rate of turning of the inflationary trajectory in the multi-field space. As a consequence, the tensor to scalar ratio of Eq.~(\ref{r-epsilon-single}) becomes $r = 16 \epsilon c_s $, implying that $r$ is now related with the field range $\Delta \phi$ along the (curved) trajectory as
\be
\Delta \phi \sim \frac{N_e}{2} \sqrt{\frac{r}{2 c_s}} . \label{Delta-phi-cs}
\ee
Current CMB constraints coming from the search of primordial non-Gaussianity give us $1 \geq c_s > 0.024 $~\cite{Ade:2015ava}. Thus, a given value of $r$ now corresponds to  a larger range $\Delta \phi$ traversed by the scalar fields, implying that the criterion 1 becomes more restrictive. On the other hand, Eq.~(\ref{epsilon-naive}) stops being valid in multi-field models of inflation (see for instance~\cite{Hetz:2016ics}). Instead, one finds:
\be
\epsilon =  \frac{1}{2} \left( \frac{\nabla V}{V} \right)^2 \left( 1 + \frac{\Omega^2}{9H^2} \right)^{-1} . \label{epsilon-Omega}
\ee
Given that $\Omega$ can be much larger than $H$, one can have a suppressed value of $\epsilon$, consistent with slow roll,  and still satisfy the second criterion. Eqs.~(\ref{Delta-phi-cs}) and (\ref{epsilon-Omega}) imply that cosmic inflation is perfectly compatible with the two string swampland criteria, particularly if $M^2 \gg \Omega^2 \gg H^2$, in which case $c_s \sim 1$ and $\epsilon \ll (\nabla V /V)^2 /2$. However, if  observations in the near future reveal the existence of primordial $B$-modes in the CMB at a level $r \gtrsim 0.01$, the two criteria would indeed be in conflict with the realization of inflation within the string landscape.

Taken together, the two criteria proposed in~\cite{Obied:2018sgi} imply  a lower bound on $\Omega / H$ --involving $\Delta$, $c$ and the number of $e$folds--  that is independent of observations. In the particular case of single field inflation the two criteria  imply
\be 
\Delta / c \gtrsim N_e .
\ee
Consequently, parameters $\Delta$ and $c$ of order 1 are already in tension with the requirement of $N_e$ of order 60 or more. 
However, in the case of multi-field inflation, $\Omega \neq 0$ can reconcile the criteria with the necessary amount of inflation because the previous bound becomes
\be
\Delta /c  \gtrsim N_e \left(1 + \frac{\Omega^2} {9 H^2} \right)^{-1/2} , \label{bound-Delta-c-Ne} \qquad {\rm or} \qquad
\frac{\Omega} { H} \geq  3 \sqrt{ \left( \frac{ c N_e} {\Delta } \right)^2 - 1 } . 
\ee
This inequality is useful to assess the characteristics that a given compactification must have in order to achieve proper inflation. The quantity $\Omega$ corresponds to the rate of turn of the inflationary trajectory, and therefore it parametrizes how non-geodesic the trajectory is in field space. Given that multi-field models derived from string compactifications are commonly characterized by non-trivial field space geometries, a misalignment between the gradient flow of the potential ($\propto V_a$) and families of geodesics of the field space is expected, inducing the appearance of a non-vanishing $\Omega$. As a consequence, if both criteria are valid, efforts to construct working models of inflation in low energy EFT derived from string theory should focus on incorporating intrinsic multi-field effects, satisfying Eq.~(\ref{bound-Delta-c-Ne}), into the task of building models.

We start this discussion in the next section where we briefly review some basic aspects of multi-field inflation. Then in Section~\ref{first_slow_roll_par} we show how  (\ref{epsilon-naive}) changes in multi-field inflation. In Section~\ref{curvature_multi-field} we briefly examine the form of the tensor to scalar ratio $r$ in multi-field models. In Section~\ref{observations-swampland} we show how the tension between observations and the requirements of $c\sim \mathcal O(1)$ and $\Delta\sim \mathcal O(1)$ relaxes in multi-field models of inflation. Finally, we offer some concluding remarks in Section~\ref{conclusions}.

\setcounter{equation}{0}
\section{Inflation in multi-field setups}
\label{inflation_mulit-field}

Let us briefly review some basic aspects about cosmic inflation in theories with more than one field. It will be enough to consider the action of Eq.~(\ref{action}) describing a system consisting of several scalar fields $\phi^a$ minimally coupled to gravity. This type of action is typical of low energy effective field theories obtained from string compactifications. For instance, in the scalar sector of ${\cal N}=1$ supergravity theories, the fields $\phi^a$ arrange into chiral fields, and the metric $G_{a b}$ is determined by a K\"ahler potential. The inflating background is described by a FRW metric of the form $ds^2 = - dt^2 + a^2(t) d {\bf x}^2$. Then, the background dynamics is determined by the following equations of motion
\bea
3 H^2 = \frac{1}{2} \dot \phi_0^2 + V , \label{Friedman} \\
D_t \dot \phi_0^a  + 3 H \dot \phi_0^a + G^{a b} V_b  = 0, \label{EOM_scalars}
\eea
where $H = \dot a / a$ is the Hubble expansion rate, and $V_b = \nabla_b V$ corresponds to a derivative of $V$ with respect to the field $\phi^b$. In the previous expression $D_t$ represents a covariant time derivative which, on a given vector $A^a$, is defined to act as $D_t A^a \equiv \dot A^a + \Gamma^a_{b c} A^b \dot \phi^c$, where $\Gamma^a_{bc}$ are the usual Christoffel symbols defined from $G_{ab}$. We have also defined $\dot \phi_0 = \sqrt{G_{a b} \dot \phi_0^a \dot \phi_0^b}$, which represents the speed of the field along the trajectory in field space. Combining the previous two equations one may further derive:
\be
\dot H = - \frac{\dot \phi_0^2}{2}  . \label{H-dot}
\ee
This equation informs us that the rate at which $H$ changes directly depends on the kinetic energy of the fields, not the potential. It is very useful to think about the multi-field dynamics in terms of an inflationary trajectory traversing the multi-field space~\cite{GrootNibbelink:2001qt}. At each point in this trajectory we may define a tangent unit-vector $T^a$ that is simply given by
\be
T^a \equiv \frac{\dot \phi_0^a}{ \dot \phi_0} .
\ee
From this definition, and with the help of the covariant derivative $D_t$ introduced earlier, we may further define a normal vector, which is given by
\be
N^a \equiv - \frac{1}{|D_t T |} D_t T^a .
\ee
Of course, we may continue defining other vectors parametrizing directions orthogonal to these two vectors, until we have a complete orthonormal basis, but this will not be needed in this discussion. Having defined these two vectors, one may introduce the rate of turning $\Omega$ of the inflationary trajectory (or simply, the angular velocity) at a given time during inflation. This quantity may be defined to satisfy the following relation:
\be
D_t T^a \equiv - \Omega N^a .
\ee
In other words, $\Omega = |D_t T |$. We have defined it in such a way that it always stay positive (unless it vanishes). We can now project Eq.~(\ref{EOM_scalars}) along the two directions $T^a$ and $N^a$ respectively. One obtains the following two equations:
\bea
\ddot \phi_0 + 3 H \dot \phi_0 + V_\phi = 0 , \label{EOM_single_scalar} \\
\Omega = \frac{V_N}{\dot \phi_0} , \label{Omega-V_N}
\eea
where $V_\phi \equiv T^a V_a$ and $V_N = N^a V_a$. The first equation has the form of the equation of motion of a single scalar field, pushed along the trajectory by the tangential component of $\nabla V$. On the other hand, the second equation gives us the value of $\Omega$ in terms of the slope of the potential along the normal direction $N^a$. It turns out that one gains nothing by projecting Eq.~(\ref{EOM_scalars}) along other directions, and that the gradient of the potential decomposes exactly as $V_a = T_a V_\phi + N_a V_N$.

\section{The first slow-roll parameter $\epsilon$} \label{first_slow_roll_par}

The first slow-roll parameter $\epsilon$ is a crucial quantity that allows us to parametrize the steady evolution of the quasi-de Sitter stage during inflation. The best definition must account for the rate of change of the Hubble parameter during inflation, that is:
\be
\epsilon \equiv - \frac{\dot H}{H^2}. \label{def-epsilon}
\ee
The requirement that the universe is in a quasi-de Sitter stage is equivalent to $\epsilon \ll 1$. Moreover, the quasi-de Sitter stage will persist as long as $\epsilon$ continues to be small for a long enough period. This is ensured by the extra requirement $|\eta| \ll 1$, where
\be
\eta \equiv \frac{\dot \epsilon}{H \epsilon} .  \label{def-eta}
\ee
Notice that both $\epsilon$ and $\eta$ in Eqs.~(\ref{def-epsilon}) and (\ref{def-eta}) are defined in terms of the time evolution of the expanding cosmological background, and not in terms of the shape of the potential $V$. However, in the treatment of multi-field models, quite often, one finds the following alternative definition:
\be
\epsilon_V \equiv \frac{1}{2} \frac{V^a V_a}{V^2} . \label{epsilon_V}
\ee
This definition has the advantage of explicitly bringing in information about the shape of the potential, via the gradient flow $\nabla_a V$. However, these two definitions are inequivalent in multi-field models of inflation, and may in fact differ substantially. The issue behind this difference is that in multi-field models the inflationary trajectory does not necessarily align with the gradient flow of the potential. To appreciate this, first notice that 
\be
\epsilon =  \frac{\dot \phi_0^2}{2 H^2},
\ee
where we have used (\ref{H-dot}) back into the definition (\ref{def-epsilon}). Then differentiating with respect to time, and using the definition of (\ref{def-eta}) together with the equation of motion (\ref{EOM_single_scalar}), one deduces that $V_\phi^2 = \frac{1}{2}  \epsilon \left[  6 H^2  - H^2  ( 2 \epsilon - \eta)   \right]^2$. Then, using $3 H^2 = \epsilon H^2 + V$, which is nothing but (\ref{Friedman}), we obtain:
\be
\frac{1}{2} \frac{ V_\phi^2 }{V^2}=  \epsilon \left[  1 + \frac{ \eta}{2 (3 - \epsilon ) }   \right]^2 .
\ee
It follows that if $\epsilon$ and $\eta$ are much smaller than $1$, then one obtains a familiar looking result:
\be
\epsilon = \frac{1}{2} \frac{ V_\phi^2 }{V^2} .
\ee
However, the right hand side of this expression does not coincide with (\ref{epsilon_V}). Here, $V_\phi$ corresponds to the projection of the gradient along the trajectory, which may be suppressed if the inflationary trajectory is not aligned with the direction of steepest descent of the potential. To quantify this misalignment we may insert $V^a = T^a V_\phi + N^a V_N$ in Eq.~(\ref{epsilon_V}). We obtain
\be
\epsilon_V = \frac{1}{2} \frac{V_\phi^2 + V_N^2}{V^2} .
\ee
Then, noticing from Eq.~(\ref{Omega-V_N}) that $V_N^2 =2 H^2 \Omega^2 \epsilon$, we finally arrive at~\cite{Hetz:2016ics}:
\be
\epsilon_V =  \epsilon \left( 1 + \frac{\Omega^2}{9H^2} \right). \label{epsilons}
\ee
Interestingly, the angular velocity $\Omega$ accounts for the misalignment between the tangent vector and the gradient flow of the potential, producing a difference between $\epsilon$ and $\epsilon_V$.\footnote{As explicitly shown in Eq.~(17) of Ref.~\cite{Hetz:2016ics}, the angular velocity $\Omega$ also produces an important difference on $\eta$ and $\eta_V \equiv \mathrm{min \,\, eigenvalue} (\nabla_a \nabla_b V) / V$. Thus, as is the case with $\epsilon_V$, any attempt to constrain $\eta_V$ from first principles, would not directly affect $\eta$ if the trajectory is non-geodesic.} Indeed, notice that $|V_N| \gg |V_\phi|$ simply means that the trajectory is aligned mostly orthogonal to the gradient flow $V_a$ of the potential. This misalignment is purely geometrical and there is no reason why it would need to be small. In particular, if the gradient flow is large and the trajectory is orthogonal to it, $\Omega^2 / H^2$ can be much larger than $1$, and one could easily have situations where both $\epsilon \ll 1$ and $\epsilon_V \sim \mathcal O(1)$ coexist.

\section{Curvature perturbations in multi-field inflation} \label{curvature_multi-field}

The angular velocity $\Omega$ does not only describe how the background inflationary trajectory meanders in field space, but it also describes how the field fluctuations are coupled. The scalar fields may be expanded in terms of background and fluctuations as $\phi^a ({\bf x},t) = \phi_0^a(t) + T^a \delta \phi_{||} + N^a \sigma + \cdots$, where the elipses denote field fluctuations along directions orthogonal to $T^a$ and $N^a$. A useful frame to study the dynamics of these fluctuations is provided by the co-moving gauge, where $\delta \phi_{||} = 0$, and where one introduces a co-moving curvature perturbation $\zeta$ by perturbing the metric as $ds^2 = - dt^2 + a^2 e^{2 \zeta} d {\bf x}^2$ (where, for the sake of simplicity, we have omitted other fluctuations appearing via the lapse and shift functions). Let us simplify further this discussion by assuming a multi-field space spanned by only two scalar fields. Then, if $\Omega$ stays approximately constant and the mass $M$ of the fluctuation orthogonal to the trajectory is much larger than $H$, one finds that $\sigma$ can be integrated out~\cite{Tolley:2009fg, Achucarro:2010jv, Achucarro:2010da} (See also~\cite{Baumann:2011su, Achucarro:2012yr, Burgess:2012dz, Gwyn:2012mw, Castillo:2013sfa, Baumann:2015nta, Achucarro:2015rfa, Tong:2017iat}), leading to an effective field theory for the $\zeta$ fluctuation~\cite{Cheung:2007st} in which the effects of the heavy field $\sigma$ are encoded via a sound speed $c_s$, given by~\cite{Achucarro:2010jv}
\be
c_s = \left(1 + \frac{4 \Omega^2}{M^2} \right)^{-1/2} ,
\ee
where $M^2 = N^a N^b \nabla_a \nabla_b V - \Omega^2 + \epsilon H^2 \mathbb{R}$ is the squared mass of the $\sigma$ fluctuation\footnote{One may also define the so called entropy mass $\mu$, which is identified through $\mu^2 = M^2 + 4 \Omega^2$. The role of the entropy mass in the multi-field dynamics has been emphasised, for instance, in Refs.~\cite{Renaux-Petel:2015mga, Achucarro:2016fby} } (here $\mathbb{R} = \mathbb{R}_{a b c d} T^a N^b T^c N^d$ is the projected Riemann tensor along the two orthogonal directions). This formula is valid even for $\Omega^2 \gg M^2$, which leads to a suppressed sound speed $c_s^2 \ll 1$. Assuming $c_s$ is slowly varying  ($| dc_s /dN | \ll  c_s$), the power spectrum of the primordial curvature perturbation with a sound speed $c_s \neq 1$ is well known, and it is given by~\cite{ArmendarizPicon:1999rj}
\be
 P_{\zeta} (k) = \frac{H^2}{ 8 \pi^2 \epsilon c_s} \frac{1}{k^3} ,
\ee
where we have omitted the effects coming from the running of the background quantities, parametrized by the spectral index $n_s - 1$. On the other hand, the power spectrum for tensor modes is given by $P_{h} (k) = 2 H^2 / \pi^2 k^3$. Then, it follows that the tensor to scalar ratio $r \equiv P_{h} (k) / P_{\zeta} (k)$ is given by
\be
r = 16 \epsilon c_s . \label{r-sound-speed}
\ee
Thus, for a fixed value of $\epsilon$ (parametrizing the quasi-de Sitter geometry) the tensor to scalar ratio is further suppressed by the sound speed $c_s$, which is always equal or smaller than 1. The relation of Eq.~(\ref{r-sound-speed}) changes in multi-field models where the mass $M$ of the orthogonal field $\sigma$ is not bigger than $H$ (for instance, see Ref.~\cite{Achucarro:2016fby}). However, one continues to find $r < 16 \epsilon$ as a generic result. In this sense, we may take $r = 16 \epsilon c_s$ as a representative result from multi-field inflation, irrespective of the precise dependence of $c_s$ on $\Omega$.

\section{Swampland criteria and inflation} \label{observations-swampland}

In this section we revisit the implications of the swampland criteria for inflation, paying special attention on the role of the multi-field effects discussed in the previous sections. There are three independent statements that can be made relating the swampland criteria (involving the parameters $\Delta$ and $c$) and inflation:\\

$\bullet$ {\bf Statement 1:} \emph{There is a lower bound on the turning rate in order for inflation to be part of the landscape.}\\

Notice that because of Eq.~(\ref{H-dot}), the relation between the field range $\Delta \phi$ and the number of $e$folds is $\Delta \phi \simeq N_e \sqrt{2 \epsilon}$, just as in the case of single field inflation. Then, the first criterion continues to imply 
\be
\Delta \gtrsim  N_e \sqrt{2 \epsilon} . \label{Delta-Ne-epsilon}
\ee
On the other hand, the second criterion may be expressed in terms of $\epsilon_V$ as $\epsilon_V \geq  c^2 /2$. Then, using this relation together with Eq.~(\ref{epsilons}), it follows that
\be
\epsilon  \geq \frac{ c^2}{2} \left( 1 + \frac{\Omega^2}{9H^2} \right)^{-1} . \label{epsilon-c}
\ee
Combining this result with Eq.~(\ref{Delta-Ne-epsilon}), one obtains the following relation anticipated in the introduction:
\be
\Delta /c  \gtrsim N_e \left(1 + \frac{\Omega^2} {9 H^2} \right)^{-1/2} . \label{Delta-c-multi}
\ee
In the case of single field inflation one recovers  $\Delta /c  \gtrsim N_e$. Equation~(\ref{Delta-c-multi}) gives us the minimal turning rate required to alleviate the tension on inflation implied by values of $\Delta$ and $c$ of order 1. In other words, given a certain type of string compactification characterized by some given values $\Delta$ and $c$, the turning rate must be such that 
\be
\frac{\Omega} { H} \geq  3 \sqrt{ \left( \frac{ c N_e} {\Delta } \right)^2 - 1 } . 
\ee
For $c/\Delta \sim 1$ and $N_e = 60$, the previous relation implies  
$\Omega \gtrsim  180 H$.\\

$\bullet$ {\bf Statement 2:}  \emph{The Lyth bound on turning trajectories makes criterion 1 more restrictive. Nevertheless, current observations still allow inflation to be compatible with criterion 1.}\\

The second statement relates the first criterion and the value of the tensor to scalar ratio $r$. As we have already seen, in multi-field models one has $r = 16 \epsilon c_s$, and so the relation between $\Delta \phi$ and the observables $r$ and $c_s$ is found to be:
\be
\Delta > \Delta \phi \sim \frac{N_e}{2} \sqrt{\frac{r}{2 c_s}} . 
\ee
(In Ref.~\cite{Baumann:2011ws}, this relation is further generalized in the broader context of EFT's of inflation). As a consequence, measurements of $r$ and $c_s$ close to their current bounds would necessarily imply a large value of $\Delta \phi$ exceeding $\Delta \sim \mathcal O(1)$. For instance, if we take $r = 0.01$ and $c_s = 0.05$, one obtains $\Delta \phi \sim 9.5$. On the other hand, $c_s \sim 1$ (which is possible if $M^2 \gg \Omega^2$) and $r < 0.001$ gives $\Delta \phi < 0.67$. Thus, a value of $c_s < 1$ makes criterion 1  more restrictive, but this crucially depends on the actual value of $r$.\\

$\bullet$ {\bf Statement 3:} \emph{There is a lower bound on the turning rate in order to meet criterion 2 and CMB observations.}\\

Our third statement connects criterion 2 with $r$.  The constraint $r < 0.07$, together with $r = 16 \epsilon c_s$, implies an upper bound given by $\epsilon < 0.0044 / c_s$. As a consequence, from Eq.~(\ref{epsilon-c}) one obtains the following requirement involving $c$ (assuming that $\Omega \gg 3 H$)
\be
 \frac{\Omega^2}{H^2} \geq c^2 10^3  c_s  .
\ee
where $c_s$ is the speed of sound of the inflaton. Thus, in order to satisfy current constraints on the tensor to scalar ratio together with the second criterion, one needs $\Omega/H \gtrsim 32 \sqrt{c_s} c $. This translates into $\Omega / H \gtrsim 32 $ if 
$c_s \sim 1$.

The previous statements may be used independently to assess the consequences of the two criteria on inflation. The first statement consists of putting together both criteria, and it is independent of any observation directly involving the size of the tensor to scalar ratio $r$. It informs us about the minimal amount of turning rate necessary in order to allow both criteria to be satisfied simultaneously. On the other hand, the second and third statements relate each of the criteria with the potential future detection of $r$ or of a reduced speed of sound. In fact, it may be noticed that the third statement is less restrictive than the first statement. It follows that, given the current status of $r$, the swampland criteria may be satisfied in multi field models as long as the following hierarchy remain valid 
\be
 \Omega \gtrsim 180 H , \label{hierarchy}
\ee
independently of the relative size of $M$ and $\Omega$. It turns out that this hierarchy is not difficult to impose. In Ref.~\cite{Achucarro:2012yr} it was examined how to achieve models with a hierarchy of the form $\Omega \gg M \gg H$, and the parameters used there may be chosen in such a way to obtain the alternative situation $M \gg \Omega  \gg H$. Furthermore, in Appendix A of Ref.~\cite{Chen:2018uul} it is shown how to construct working models of two-field inflation with arbitrary values of $\Omega$ and $M$ (and large values of the gradient flow $\nabla V$).

Before concluding, let us digress to highlight an important caveat that could alter the validity of statement 2. Strictly speaking, the Lyth bound refers to the traversed distance covered by the non-geodesic trajectory in field space, which may differ significantly from the geodesic distance between any two points on the trajectory (bounded by the first swampland criterion). This implies that, in practice, our statement 2 might be weaker. That is, it is perfectly conceivable that a measurement of $r$ close to current bounds could be compatible with non-geodesic trajectories that do not violate the first criterion. To assess this, one would have to examine the resulting non-geodesic trajectories of specific models, and compare the maximum geodesic distance against the maximum range $\Delta$.

\section{Conclusions}\label{conclusions}

For now, the swampland criteria must be regarded as speculative properties of the landscape, and more study is in order to evaluate their validity and potential consequences for low energy string compactifications\footnote{For instance, over the last years, progress~\cite{Kallosh:2014wsa, Ferrara:2014kva, Bergshoeff:2015jxa, Polchinski:2015bea, Aalsma:2018pll} has been made in understanding the stabilization of moduli within KKLT-like constructions~\cite{Kachru:2003aw}. If such constructions turn out to be realized in string theory, they would render the second criterion invalid.}. At any rate, the connection between the swampland criteria and inflation is more subtle in the case of inflationary cosmology with more than one field. As we have seen, when intrinsically multi-field effects are taken into account, the first criterion becomes more restrictive whereas the second criterion is relaxed. These multi-field effects are parametrized by the rate of turning $\Omega$ of the inflationary trajectory, which has a number of other interesting phenomenological consequences for early universe cosmology. Crucially, these effects cannot be ignored in low energy EFT's derived from inflation, which are generically inhabited by a large number of scalar fields and, at the same time, are characterized by having a non-trivial field space geometry. Put differently: the swampland criteria can lead to a wrong assessment on inflation if we apply them on naive (canonical) single field constructions that only represent a small class of models within the swampland. 

Although inflation is still compatible with both, the swampland criteria and observations, we have confirmed that a near future observation of tensor modes would indeed put pressure on the realization of inflation within string theory. Coming experiments will be able to either uncover or place constraints on the presence of tensor modes up to $r \sim 0.001$. As already pointed out in the literature, a detection of tensor modes about that value would require large field excursions of the inflaton measured along the trajectory, which could be in conflict with criterion 1.\footnote{But, as discussed at the end of the previous section, this would have to be examined model by model if their trajectory is non-geodesic.} However, as we have stressed here, it would not necessarily be in conflict with criterion 2.

\subsection*{Acknowledgements}
We would like to thank Thomas Grimm, Renata Kallosh, Andrei Linde, Miguel Montero, Spyros Sypsas, Irene Valenzuela and Yvette Welling for discussions on the swampland criteria constraints. AA acknowledges support by the Netherlands Organization for Scientific Research (NWO), the Netherlands  Organization for Fundamental Research in Matter (FOM), the Basque Government (IT-979-16) and the Spanish Ministry MINECO (FPA2015-64041-C2-1P). GAP acknowledges support from the Fondecyt Regular project number 1171811.

\end{document}